\documentclass[twocolumn]{aastex631}
\shorttitle{JWST probes star clusters at z=4}
\shortauthors{Vanzella et al.}

\usepackage{hyperref}  
\hypersetup{colorlinks=true,linkcolor=[rgb]{1.,0.2,0.2},citecolor=[rgb]{0.1,0.4,1.},filecolor=[rgb]{0.7,0.2,0.2},urlcolor=[rgb]{0.7,0.2,0.2}}

\usepackage{color}
\definecolor{blue}{rgb}{0., 0., 1}

\newcommand {\T}{Table\,}

\newcommand{\nvalone}{\textrm{N}\textsc{v}}
\newcommand{\civalone}{\textrm{C}\textsc{iv}}

\newcommand{\ha}{\ifmmode {\rm H}\alpha \else H$\alpha$\fi}
\newcommand{\hb}{\ifmmode {\rm H}\beta \else H$\beta$\fi}
\newcommand{\lya}{\ifmmode {\rm Ly}\alpha \else Ly$\alpha$\fi}
\newcommand{\pg}{\ifmmode {\rm P}\gamma \else Pa$\gamma$\fi}
\newcommand{\lyb}{\ifmmode {\rm Ly}\beta \else Ly$\beta$\fi}
\newcommand{\lyg}{\ifmmode {\rm Ly}\gamma \else Ly$\gamma$\fi}

\newcommand{\civ}{\textrm{C}\textsc{iv}\ensuremath{\lambda1548,1550}}

\newcommand{\heii}{\textrm{He}\textsc{ii}\ensuremath{\lambda1640}}
\newcommand{\oiiiuv}{\textrm{O}\textsc{iii}]\ensuremath{\lambda1661,1666}}

\newcommand{\flyc}{\ifmmode  \mathrm{f}_\mathrm{esc}\mathrm{(LyC)} \else $\mathrm{f}_\mathrm{esc}\mathrm{(LyC)}$\fi}

\def\ergs{\ifmmode \mathrm{erg\hspace{1mm}s}^{-1} \else erg s$^{-1}$\fi}

\def\micron{\ifmmode \mu\mathrm{m} \else $\mu$m\fi}
\def\msun{\ifmmode \mathrm{M}_{\odot} \else M$_{\odot}$\fi}
\def\msunyr{\ifmmode \mathrm{M}_{\odot} \hspace{1mm}{\rm yr}^{-1} \else $\mathrm{M}_{\odot}$ yr$^{-1}$\fi}
\def\zsun{\ifmmode Z_{\odot} \else Z$_{\odot}$\fi}
\def\lsun{\ifmmode L_{\odot} \else L$_{\odot}$\fi}
\def\mstar{\ifmmode \mathrm{M}_{\star} \else M$_{\star}$\fi}

\begin{document}


\title{Early results from GLASS-JWST. VII: evidence for lensed, gravitationally bound proto-globular clusters at z=4 in the Hubble Frontier Field A2744\footnote{Based on observations collected with JWST under the ERS program 1324 (PI T. Treu)}}

\email{eros.vanzella@inaf.it}

\author[0000-0002-5057-135X]{E.~Vanzella}
\affiliation{INAF -- OAS, Osservatorio di Astrofisica e Scienza dello Spazio di Bologna, via Gobetti 93/3, I-40129 Bologna, Italy}

\author[0000-0001-9875-8263]{M.~Castellano}
\affiliation{INAF Osservatorio Astronomico di Roma, Via Frascati 33, 00078 Monteporzio Catone, Rome, Italy}

\author[0000-0003-1383-9414]{P.~Bergamini}
\affiliation{Dipartimento di Fisica, Università degli Studi di Milano, Via Celoria 16, I-20133 Milano, Italy}
\affiliation{INAF - OAS, Osservatorio di Astrofisica e Scienza dello Spazio di Bologna, via Gobetti 93/3, I-40129 Bologna, Italy}

\author[0000-0002-8460-0390]{T.~Treu}
\affiliation{Department of Physics and Astronomy, University of California, Los Angeles, 430 Portola Plaza, Los Angeles, CA 90095, USA}

\author[0000-0001-9261-7849]{A.~Mercurio}
\affiliation{INAF – Osservatorio Astronomico di Capodimonte, Via Moiariello 16, I-80131 Napoli, Italy}

\author[0000-0002-9136-8876]{C.~Scarlata}\affiliation{
School of Physics and Astronomy, University of Minnesota, Minneapolis, MN, 55455, USA}

\author[0000-0002-6813-0632]{P.~Rosati}
\affiliation{Dipartimento di Fisica e Scienze della Terra, Università degli Studi di Ferrara, Via Saragat 1, I-44122 Ferrara, Italy}
\affiliation{INAF - OAS, Osservatorio di Astrofisica e Scienza dello Spazio di Bologna, via Gobetti 93/3, I-40129 Bologna, Italy}

\author[0000-0002-5926-7143]{C.~Grillo}
\affiliation{Dipartimento di Fisica, Università degli Studi di Milano, Via Celoria 16, I-20133 Milano, Italy}
\affiliation{INAF - IASF Milano, via A. Corti 12, I-20133 Milano, Italy}

\author[0000-0003-3108-9039]{A.~Acebron}
\affiliation{Dipartimento di Fisica, Università degli Studi di Milano, Via Celoria 16, I-20133 Milano, Italy}
\affiliation{INAF - IASF Milano, via A. Corti 12, I-20133 Milano, Italy}

\author[0000-0001-6052-3274]{G.~B.~Caminha}
\affiliation{Max-Planck-Institut f\"ur Astrophysik, Karl-Schwarzschild-Str. 1, D-85748 Garching, Germany}

\author[0000-0001-6342-9662]{M.~Nonino}
\affiliation{INAF Osservatorio Astronomico di Trieste, Via Tiepolo  11, 34123, Trieste, Italy}

\author[0000-0003-2804-0648 ]{T.~Nanayakkara}
\affiliation{Centre for Astrophysics and Supercomputing, Swinburne University of Technology, PO Box 218, Hawthorn, VIC 3122, Australia}
\affiliation{ARC Centre of Excellence for All Sky Astrophysics in 3 Dimensions (ASTRO 3D), Australia)}

\author[0000-0002-4140-1367]{G.~Roberts-Borsani}
\affiliation{Department of Physics and Astronomy, University of California, Los Angeles, 430 Portola Plaza, Los Angeles, CA 90095, USA}

\author[0000-0001-5984-0395]{M.~Bradac}
\affiliation{
University of Ljubljana, Department of Mathematics and Physics, Jadranska ulica 19, SI-1000 Ljubljana, Slovenia}
\affiliation{
Department of Physics and Astronomy, University of California Davis, 1 Shields Avenue, Davis, CA 95616, USA}

\author[0000-0002-9373-3865]{X.~Wang}
\affiliation{Infrared Processing and Analysis Center, Caltech, 1200 E. California Blvd., Pasadena, CA 91125, USA}

\author[0000-0003-2680-005X]{G.~Brammer}
\affiliation{Cosmic Dawn Center (DAWN), Denmark}
\affiliation{Niels Bohr Institute, University of Copenhagen, Jagtvej 128, DK-2200 Copenhagen N, Denmark}

\author[0000-0002-6338-7295]{V.~Strait}
\affiliation{Cosmic Dawn Center (DAWN), Denmark}
\affiliation{Niels Bohr Institute, University of Copenhagen, Jagtvej 128, DK-2200 Copenhagen N, Denmark}

\author[0000-0003-0980-1499]{B.~Vulcani}
\affiliation{INAF Osservatorio Astronomico di Padova, vicolo dell'Osservatorio 5, 35122 Padova, Italy}

\author[0000-0002-0441-8629]{U.~Mestric}
\affiliation{INAF -- OAS, Osservatorio di Astrofisica e Scienza dello Spazio di Bologna, via Gobetti 93/3, I-40129 Bologna, Italy}

\author[0000-0003-1225-7084]{M.~Meneghetti}
\affiliation{INAF -- OAS, Osservatorio di Astrofisica e Scienza dello Spazio di Bologna, via Gobetti 93/3, I-40129 Bologna, Italy}

\author[0000-0002-6175-0871]{F.~Calura}
\affiliation{INAF -- OAS, Osservatorio di Astrofisica e Scienza dello Spazio di Bologna, via Gobetti 93/3, I-40129 Bologna, Italy}

\author[0000-0002-6586-4446 ]{Alaina Henry}
\affiliation{Space Telescope Science Institute, 3700 San Martin
  Drive, Baltimore, MD, 21218, USA}
  \affiliation{Center for Astrophysical Sciences, Department of Physics \& Astronomy, Johns Hopkins University, Baltimore, MD 21218, USA}

\author[0000-0001-8600-7008]{A.~Zanella}
\affiliation{INAF Osservatorio Astronomico di Padova, vicolo dell'Osservatorio 5, 35122 Padova, Italy}

\author[0000-0001-9391-305X]{M.~Trenti}
\affiliation{School of Physics, University of Melbourne, Parkville 3010, VIC, Australia}
\affiliation{ARC Centre of Excellence for All Sky Astrophysics in 3 Dimensions (ASTRO 3D), Australia}

\author[0000-0003-4109-304X]{K.~Boyett}
\affiliation{School of Physics, University of Melbourne, Parkville 3010, VIC, Australia}
\affiliation{ARC Centre of Excellence for All Sky Astrophysics in 3 Dimensions (ASTRO 3D), Australia}

\author[0000-0002-8512-1404]{T.~Morishita}
\affiliation{Infrared Processing and Analysis Center, Caltech, 1200 E. California Blvd., Pasadena, CA 91125, USA}

\author[0000-0003-2536-1614]{A.~Calabr\`o}
\affiliation{INAF Osservatorio Astronomico di Roma, Via Frascati 33, 00078 Monteporzio Catone, Rome, Italy}

\author[0000-0002-3254-9044]{K.~Glazebrook}
\affiliation{Centre for Astrophysics and Supercomputing, Swinburne University of Technology, PO Box 218, Hawthorn, VIC 3122, Australia}

\author[0000-0001-9002-3502]{D.~Marchesini}
\affiliation{Department of Physics and Astronomy, Tufts University, 574 Boston Ave., Medford, MA 02155, USA}

\author[0000-0003-3195-5507]{S.~Birrer}
\affiliation{Kavli Institute for Particle Astrophysics and Cosmology and Department of Physics, Stanford University, Stanford, CA 94305, USA}
\affiliation{SLAC National Accelerator Laboratory, Menlo Park, CA, 94025}
\affiliation{Department of Physics and Astronomy, Stony Brook University, Stony Brook, NY 11794, USA}

\author[0000-0002-8434-880X]{L.~Yang}
\affiliation{Kavli Institute for the Physics and Mathematics of the Universe, The University of Tokyo, Kashiwa, Japan 277-8583}

\author[0000-0001-5860-3419]{T. Jones}
\affiliation{Department of Physics and Astronomy, University of California Davis, 1 Shields Avenue, Davis, CA 95616, USA}


%
%
%




\begin{abstract}

We investigate the blue and optical rest-frame sizes ($\lambda \simeq 2300-4000$\AA) of three compact star-forming regions in a galaxy at z=4 strongly lensed ($\times30, \times45, \times100$) by the Hubble Frontier Field galaxy cluster A2744 using GLASS-ERS JWST/NIRISS imaging at $1.15\mu m$, $1.50\mu m$ and $2.0\mu m$ with PSF $\lesssim 0.1''$. In particular, the Balmer break is probed in detail for all multiply-imaged sources of the system.
With ages of a few tens of Myr, stellar masses in the range $(0.7-4.0)\times10^{6}$~\msun\ and optical/ultraviolet effective radii spanning the interval $3<\rm R_{\tt eff} < 20$~pc, such objects are currently the highest redshift (spectroscopically-confirmed) gravitationally-bound young massive star clusters (YMCs), with stellar mass surface densities resembling those of local globular clusters. Optical (4000\AA, JWST-based) and ultraviolet (1600\AA, HST-based) sizes are fully compatible.
The contribution to the ultraviolet underlying continuum emission ($1600$\AA) is $\sim 30$\%, which decreases by a factor of two in the optical for two of the YMCs ($\sim 4000$\AA~rest-frame), reflecting the young ages ($<30$ Myr) inferred from the SED fitting and supported by the presence of high-ionization lines secured with VLT/MUSE.
Such bursty forming regions enhance the sSFR of the galaxy, which is $\simeq 10$~Gyr$^{-1}$. 
This galaxy would be among the extreme analogs observed in the local Universe having high star formation rate surface density and high occurrence of massive stellar clusters in formation. 
\end{abstract}



\section{Introduction} \label{sec:intro}

The hierarchical nature of star formation emerges whenever the available spatial resolution increases. This is evident from studies of local Universe starburst galaxies \citep[e.g.,][]{adamo20, Mehta21,calzetti15} and at cosmological distance, $z>1$, where {\it clumps} of sub-kpc size typically populate star-forming galaxies \citep[][]{Elemgreen2009, Genzel2011, livermore15, Iani2021MNRAS, zanella15, zanella19, Guo2018, sok2022}.  
Such clumps, when observed through gravitational lensing amplification, break down into smaller star-forming complexes, each of a $\lesssim 100$~pc scale  \citep[e.g.][]{rigby17, johnson17, cava18, mirka19,vanz_mdlf,  Mestric22}.  When magnification is large enough to resolve them, clumps break down into single stellar clusters, with sizes smaller than a few tens of parsecs, as exemplified by the exceptional {\tt Sunburst} galaxy at z=2.37
\citep[e.g.][]{vanz_sunburst,vanz22_CFE} or the candidate globular cluster precursors identified at $z\sim6$ \citep[e.g., by][]{vanz_paving, vanz19, Welch22_clumps}. 

For reasons summarized below, the ultimate goal in the study of the star-forming modes  is the identification of single star clusters  at all epochs (or at least the most massive gravitationally-bound ones, with stellar mass $M_*< 10^{5-7}$\msun). Meeting this goal requires accessing angular scales smaller than $3.0(4.5)$ milli-arcsec, roughly corresponding to physical scales smaller than 25~pc at $z$=2(6). Young massive star clusters (YMCs) are key sources of mechanical and radiative feedback and efficient producers of ionizing radiation \citep[e.g.,][]{heckman11, bik18}
and eventually key actors in shaping and modulating the star formation activity of the host galaxy \citep[][]{sirressi22_HARO11,james16, calzetti15_ngc5253}. Star clusters might also play a role in regulating the leakage of ionizing radiation by carving ionized channels in the interstellar medium \citep[ISM,][and references therein]{rivera19,vanz_sunburst,vanz22_CFE}, making them additional contributors to cosmic reionization \citep[e.g.,][]{he20_ricotti}. 

Currently, the only way to access such tiny physical scales at high redshift is to exploit gravitational lensing at large magnification $\mu$ ($\mu > 20$). However, given the small cosmic volume probed so far in the high magnification regime (which scales as A($>\mu_o$)~$\sim \mu_o^{-2}$), only a few cases are currently known \citep[e.g.,][and references therein]{rigby17, Welch22_clumps, vanz22_CFE}. 
A common limitation in such studies is the difficulty to access optical rest-frame wavelengths at high angular resolution, i.e. $\lesssim 0.1''$ FWHM. Up to now, and in absence of assistance from strong lensing magnification, this resolution can only be achieved with the Hubble Space Telescope (HST) in the blue bands, probing rest-frame UV only. 

In this work, we report on multi band James Webb Space Telescope (JWST)/NIRISS imaging and spectroscopy \citep[][]{NIRISS} of an exceptionally magnified system, at $z$=4. The observations in the F115W, F150W and F200W filters 
provide, for the first time,  detailed constraints in the optical rest-frame of the three massive star clusters identified in the galaxy.

We assume a flat cosmology with $\Omega_{M}$= 0.3,
$\Omega_{\Lambda}$= 0.7 and $H_{0} = 70$ km s$^{-1}$ Mpc$^{-1}$. 
All magnitudes are given in the AB system \citep{Oke_1983}.

\begin{figure*}
\center
 \includegraphics[width=\textwidth]{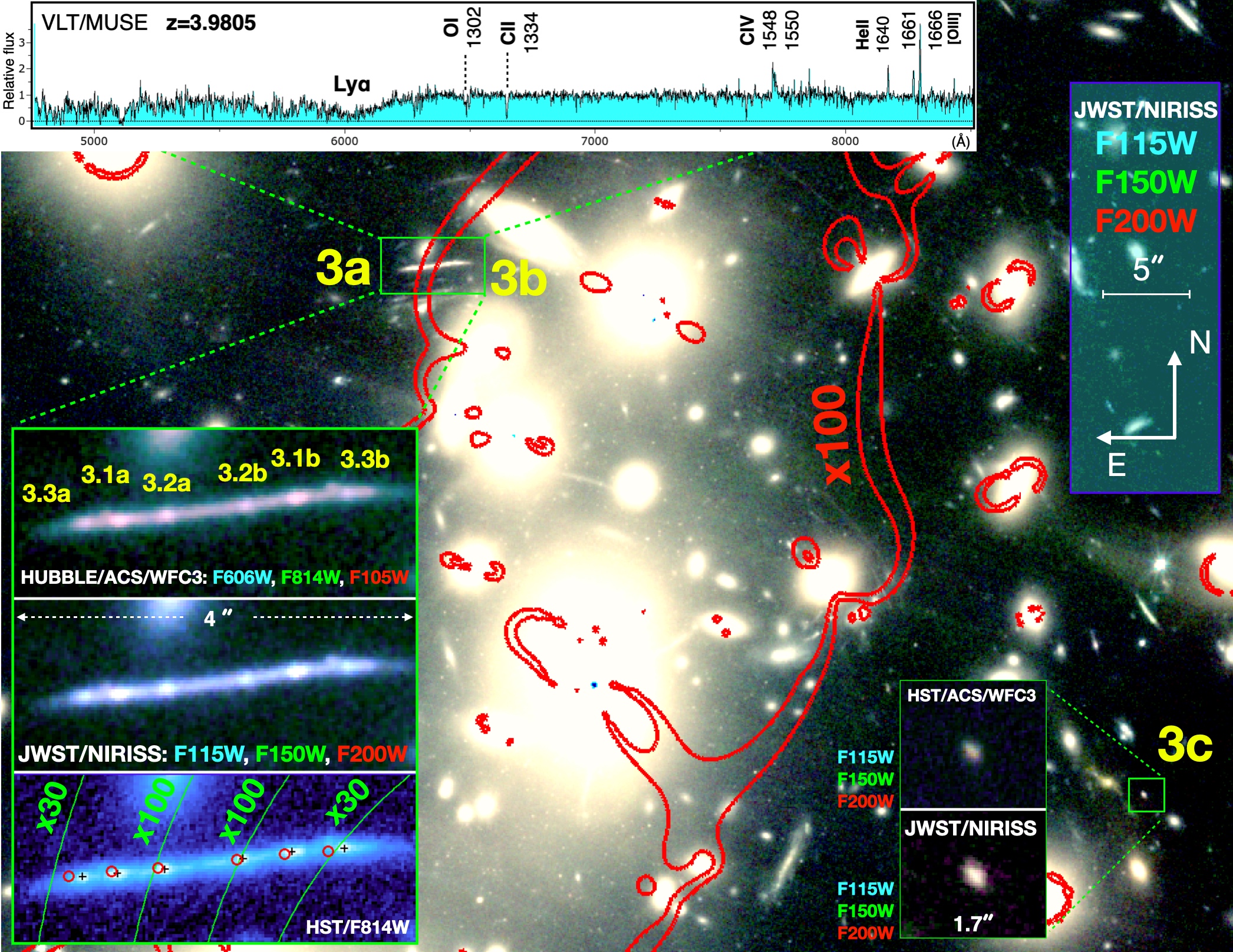}
 \caption{Color composition of JWST/NIRISS imaging of a portion of the cluster A2744 (the bands used for the red, green and blue channels are indicated in the top-right). The highly magnified arclet is marked with a green rectangle in the main panel, along with its zoomed version (lower-left insets) where the compact star clusters, 3.1a, 3.2a, 3.3a and 3.1b, 3.2b, 3.3b are shown.
 The two red contours mark the locus for magnification $\times 100$. Lower-left inset, from top to bottom: the color image based on HST bands, the JWST/NIRISS color image, and the HST/F814W image with the green lines marking the locations at magnification $\mu_{TOT} = 30$ and 100; black crosses indicate the observed position of the knots and the red circles the predicted position from the lens model \citep[][]{bergamini22_a2744}. Bottom-right inset: the same color images are shown for the counter-image 3c. Top-inset: the VLT/MUSE spectrum of the arclet reporting some of the relevant absorption and emission lines.}
 \label{fig:pano}
\end{figure*}

\section{JWST/NIRISS imaging and spectroscopy of HFF~A2744} \label{sec:data}

JWST/NIRISS $2.2'\times2.2'$ observations centered on the Hubble Frontier Field~A2744 \citep[HFFs,][]{Lotz_2017HFF}
were acquired on June 28-29 2022. The cluster was observed with the wide field slitless spectroscopy (WFSS) mode, that provides an average spectral resolution of R=150. The spectra and associated direct imaging were obtained using  the F115W, F150W and F200W blocking filters, providing a full spectral coverage between 1.0 and 2.2$\mu$m. The observations have an  
exquisite PSF of $\simeq 0.08'' , 0.10'', 0.11''$ from blue to red (as measured from four non-saturated stars in the field). Imaging integration time was 2830s while 10400s were split evenly between the two orthogonal GR150R and GR150C grisms.
Details on the data acquisition, observing strategy and data reduction are described by \citet{TreuGlass22}  and \citet{Guido2022}. 
NIRISS pre-imaging has been projected on a final pixel scale of $0.03''$, to mitigate the effects of the sharp PSF at the observed wavelengths and to match the pixel scale  of existing HST images. 
In addition to the new JWST/NIRISS data, we also make use of archival VLT/MUSE data \citep[][]{mahler18, richard2021, bergamini22_a2744} and Hubble Frontier Field imaging \citep[][]{Lotz_2017HFF}. 

\begin{figure*}
\center
 \includegraphics[width=\textwidth]{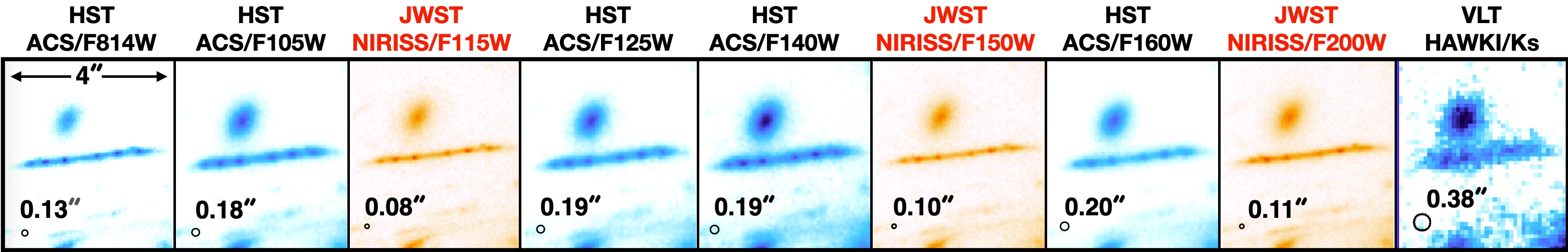}
 \caption{Cutouts of the arclet as observed with HST, JWST and VLT filters. Small circles in the lower-left corner indicate the PSF size in each instrument/band (for Hubble and VLT they are taken from \citet{merlin16}, whereas for JWST the PSFs are estimated with available non-saturated stars).  }
 \label{fig:cutouts}
\end{figure*}

\section{A super-magnified source at z=4}

\subsection{The ultraviolet view: HST and VLT/MUSE}
We focus here on ``System 3'', a highly magnified source confirmed at z=3.9805 and observed with VLT/MUSE by \citet{mahler18}. System 3 is split into 
three multiple images (3a, 3b and 3c in Figure~\ref{fig:pano}, see also \citealt{bergamini22_a2744}). 
Images 3a,b straddle the critical line, which produces  
large magnification values resulting in a bright arclet with a total magnitude F814W~$=23.2\pm0.02$ (at $\lambda \simeq 1600$\AA~rest-frame). Based on the recent high-precision lens model described in \citet{bergamini22_a2744}, the arclet is split into two images which lie on both sides of the critical line at z=3.9805. In each image, (at least) 
three additional compact star-forming regions appear mirrored on 
each side and are subjected to total magnification ($\mu_{Tot}$) values spanning the interval $30\times -100\times$, with a large associated tangential stretch ($\mu_{Tang}$) as reported in the Table~\ref{tab:summary} (see also Figure~\ref{fig:pano}). Such a mirrored triplet of star-forming regions is evident in all space-based bands obtained with HST and JWST (Figure~\ref{fig:cutouts}). 
The rest-frame ultraviolet high signal to noise ratio VLT/MUSE spectrum of the arclet shows absorption lines and high-ionization nebular emission lines in emission (\civ, \heii, \oiiiuv), along with the absence of \lya. 
The third least magnified counter-image (3c) is too faint and not confirmed with VLT/MUSE spectroscopy. 

{\tt GALFIT} fitting \citep[][]{Peng_2010} of each star-forming knot was performed in the HST F814W image following the usual methodology described in, e.g., \citet{vanz22_CFE}.  
Residual images were visually inspected and checked so that the remaining underlying median signal was of the same level as the median value calculated in the ``intra-knot'' regions. In this way, only the signal from the clean arclet, free from the nucleated knots, is left after subtracting the {\tt GALFIT} models (see Figure~\ref{fig:galfit}). We adopt a Sersic index $n=1$ (and note that equivalent results are obtained with Gaussian profiles). Only the multiple imaged knot 3.1 (mirrored a,b) appears elongated along the tangential shear with an effective radius R$_{\rm eff}=2.0 \pm 1.0$ pixel, whereas the others appear barely resolved with an R$_{\rm eff}= 1_{-0.5}^{+1.0}$ pixel.  
The inferred total magnitudes span the range $26.5-26.8$,
with a typical uncertainty of 0.3 mags
(and corresponding to 30.5-31.5 intrinsic). The sizes and absolute magnitudes in the HST/F814W band are reported in Table~\ref{tab:summary}, while Figure~\ref{fig:galfit} shows the modeled knots and the residual diffuse arclet after subtraction.

\subsection{A sharp rest-frame optical view: JWST/NIRISS}

\begin{figure*}[h]
\center
 \includegraphics[width=\textwidth]{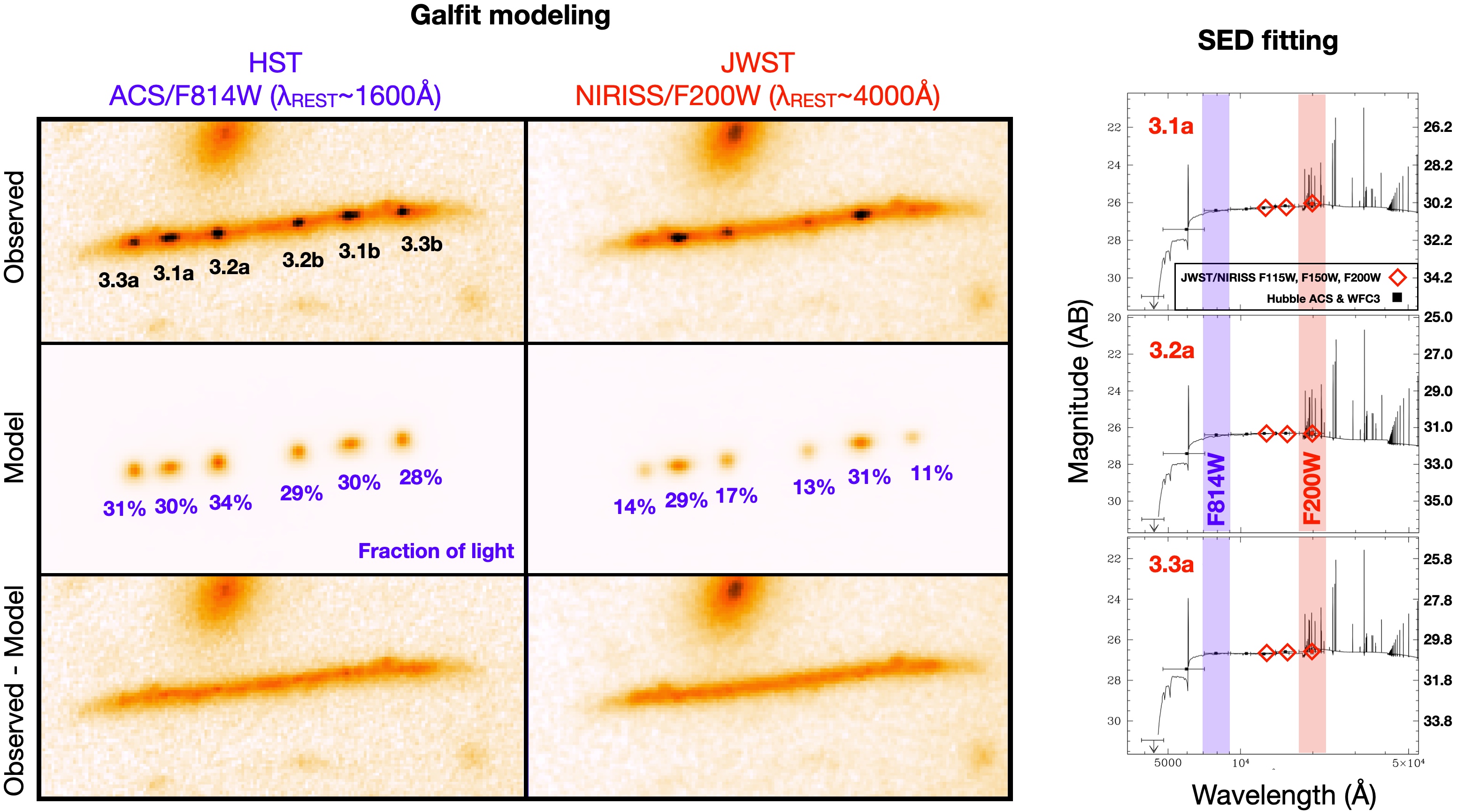} 
 \caption{{\it Left panel}: {\tt GALFIT} modeling of the compact knots in the images 3a and 3b performed in the HST/ACS F814W (left column) and JWST/NIRISS F200W (right column) bands. The fraction of light the compact sources contribute to the total underlying region is reported in blue in the middle panels (fluxes are measured within circular apertures of $2\times FWHM_{PSF}$). {\it Right panel}: SED fitting is shown for three of the knots (best fit parameters are reported in Table~\ref{tab:summary}). The F814W and F200W bands are indicated with vertical transparent stripes; magnitudes on the left(right) Y-axis are observed(intrinsic).}
 \label{fig:galfit}
\end{figure*}

The three JWST/NIRISS filters used for pre-imaging probe the 2300\AA, 3000\AA~and 4000\AA~rest-frame. Although these wavelengths were partially covered by the HST WFC3/F125W and F140W/F160W filters, the new JWST/NIRISS images have an angular resolution about two times sharper than HST. The PSF was
extracted from non-saturated stars in the same field of view and used in the {\tt GALFIT} runs. JWST/NIRISS imaging produces nearly diffraction-limited images at the observed wavelengths which are undersampled in the raw data \citep[][]{Guido2022}.
The final PSF FWHM are $0.08''$, $0.10''$ and $0.11''$ in the F115W, F150W and F200W, respectively (see Figure~\ref{fig:cutouts}).
F200W (K-band), instead, is unique at this stage and probes spatial scales four times sharper than the ground-based VLT/HAWKI \citep[][]{patig2008}. Furthermore, it is at least 2.5 magnitudes deeper. We show the F200W image  (4000\AA~rest-frame) of the arclet in Figure~\ref{fig:galfit}, where it appears clear that the optical sizes are not dissimilar from the ultraviolet ones. Size and morphology were determined with 
{\tt GALFIT} following the same procedure applied to the HST data. Results are listed in Table~\ref{tab:summary}. 

\subsection{Very large magnification}

The predicted image 3c 
can be used to probe 
the magnification the system is subject to by looking at the relative magnifications among multiple images \citep[e.g.,][]{vanz_paving}. In particular, the flux ratio f(3[a+b])/f(3c) measured in the F814W band is $19\pm 1$ which sets a stringent lower limit to the magnification of 3(a+b): 
\begin{equation}
\mu(3[a+b]) > 19\times \mu(3c)>73,
\end{equation}
where  $\mu(3c)=3.8\pm0.2$ is well constrained from the lens model \citep[][]{bergamini22_a2744}. Such a lower limit is due to the fact that only a portion of the galaxy is probed at positions 3a (or 3b) and that the magnification gradient is not included in this estimate. The inferred value is in line with a very high magnification regime. The lens model predictions therefore accurately reproduce the magnification and positions of all multiple images/knots (see Figure~\ref{fig:pano}). The total magnifications ($\mu_{Tot}$) derived at the predicted positions are $\mu_{Tot}\simeq 30, 45$ and 100 for images 3.3(a,b), 3.1(a,b) and 3.2(a,b), respectively. These values are dominated by the tangential magnification ($\mu_{Tang}$), such that $\mu_{Tang}\simeq 20, 30$ and 60 with statistical uncertainties not larger than 20\% at two sigma.
The de-lensed effective radii (parsec) in the ultraviolet (HST-based) and optical (JWST-based) wavelengths are reported in Table~\ref{tab:summary} and are fully consistent with each other within uncertainties.

\subsection{Physical properties of the star forming knots}

Physical quantities of each star forming knot of the arclet have been derived with Spectral Energy Distribution (SED) fitting, along with the host galaxy probed by image 3c. 
The extraction of the HST and JWST photometry for this specific source is  described in Appendix~\ref{appA}.
We use standard BC03 \citep[][]{Bruzual_2003} templates and 
adopt the following priors: Chabrier initial mass function, exponentially
declining star-formation histories with e-folding times of
$0.1<\tau<15.0$~Gyr, and
the extinction laws from both Small
Magellanic Cloud \citep[][]{prevot84} and \citet{Calzetti_2000}. We considered the following
range of physical parameters: $0.0< E(B-V) < 1.1$,
Age~$>$~1Myr (defined as the onset of the star-formation episode)
and metallicity Z/Z$_{\odot}$ = 0.02, 0.2, 1.0.
The best-fit de-lensed values are reported in Table~\ref{tab:summary}. The three nucleated star-forming regions show a rather flat ultraviolet slope ($\beta \simeq -1.9$, $F_{\lambda} \sim \lambda^{\beta}$), 
stellar masses in the range $\simeq (1-7)\times 10^{6}$~\msun, ages in the range $20-70$ Myr and SFR of a few percent \msunyr, with E(B-V)~$\simeq 0.1$. The entire galaxy hosting the three compact star forming regions is probed by image 3c and its properties are also reported in Table~\ref{tab:summary}. 

\section{Young massive stellar clusters at $z=4$}

Despite the large magnification stretch, the very compact observed sizes of the three stellar systems imply they are truly small, and eventually having sizes comparable to those of young massive star clusters in the local Universe. 
To explore if they are gravitationally-bound star clusters we derive the dynamical age $\Pi$ (which is the ratio between the age of the system and crossing time T$_{\rm CR}$, \citealt{gieles11}). This quantity has been extensively used to identify bound star clusters in the local Universe \citep[e.g.,][and recently for the first time at cosmological distance, \citealt{vanz22_CFE}]{adamo20extreme,ryon17}.

In particular, 
the crossing time expressed in Myr is defined as ${\rm T_{CR} = 10 \times (R_{\tt eff}^{3} / GM)^{0.5}}$, where M and ${\rm R_{eff}}$ are the stellar mass and the effective radius, respectively, and ${\rm G \approx 0.0045~pc^{-3} \msun^{-1} Myr^{-2}}$ is the gravitational constant. Stellar systems evolved for more than a crossing time have ${\rm \Pi > 1}$, suggestive of being bound.
In calculating $\Pi$ we assume that all the involved quantities have an uncertainty of 50\%, except the magnification for which we take 30\% (corresponding to 3-sigma statistical error, \citealt{bergamini22_a2744}). 
Sources 3.1, 3.2 and 3.3 (``a'' or ``b'' groups) have $\Pi>1$ which implies they are gravitationally-bound objects. Given their extremely small sizes (R$_{\rm eff}<$20~pc), they likely are the highest redshift, spectroscopically confirmed, stellar clusters probed so far.
The specific star formation rates (sSFR) are significantly large 
$>10$~Gyr$^{-1}$, along with their stellar mass surface densities which resemble those ones observed in local globular (or young massive) clusters (Table~\ref{tab:summary}).
Such objects result to be relatively young ($20-80$ Myr old), though not as young as other cases showing spectral signatures like p-cygni profiles of \nvalone\ and/or \civalone\ and/or very blue colors indicating $<10$~Myr stellar populations \citep[][]{vanz22_CFE}. Rather, their properties resemble those of another lensed physical pair of stellar systems (likely massive star clusters) at z=3.222 which shows a similar flat ultraviolet slope ($\beta$), absence of \lya\ emission, the presence of high-ionization nebular lines and ages of the same order of magnitude as those derived here \citep[$>20$ Myr,][Appendix C; see also the SED fits in \citealt{vanz_id14}]{vanz_mdlf}.

By means of the JWST/NIRISS resolution in the K-band (NIRISS/F200W), we can compare the fraction of optical light (4000\AA~rest-frame) the star clusters contribute to the underlying host region to the corresponding fraction measured in the ultraviolet (1600\AA~rest-frame, HST/F814W). Two out of the three stellar clusters (3.3 and 3.2) are more prominent in the UV (contributing to 30\% of the light inferred within the adopted circular aperture of $2\times FHWM$) and significantly decrease to $\sim 12\%$ in the F200W band. Instead, in the case of system 3.1, such fraction of light remains constant in both bands (Figure~\ref{fig:galfit}). Such behaviors are consistent with the inferred young ages for 3.2 and 3.3 (decreasing flux) and with the relatively older age for 3.1 (constant flux). Similar results have been recently observed in the local starburst galaxy Haro11 which hosts a multitude of massive stellar clusters \citep[][]{sirressi22_HARO11}.

Finally, the host galaxy, as probed by image 3c, has a relatively low stellar mass, $2\times 10^{8}$~\msun, and a high star formation rate surface density ($\Sigma_{SFR}$, $M_{\odot}$~yr$^{-1}$~kpc$^{-2}$) log$_{10}(\Sigma_{SFR})=0.6$, for which a high occurrence of forming star clusters is expected from studies of local analog dense starburst galaxies, showing large star cluster formation efficiencies \citep[][]{adamo20extreme, adamo20}.

\section{Summary and conclusions}
In this work, we study a highly magnified galaxy at z=4, straddling the critical line of the Hubble Frontier Field cluster A2744. Several star forming regions are revealed as very compact objects, despite the large tangential stretch provided by gravitational lensing. In particular, the new JWST/NIRISS imaging of the system allows us to:

\noindent $\bullet$ probe the star forming knots at unprecedented angular resolution in the near-infrared (JWST/NIRISS F200W, corresponding to rest-frame optical), matching the existing high-resolution rest-frame ultraviolet view from HST/F814W. The identified nucleated forming knots show comparable effective radii in the ultraviolet and in the optical, spanning the range $3-20$ pc.

\noindent $\bullet$ derive SED fitting locally for each compact region taking advantage of the wavelength coverage of JWST/NIRISS that encompasses the rest-frame Balmer-break region at $z=4$.

Such parsec-scale sources are inferred to be a few tens of Myr old gravitationally-bound stellar clusters, with estimated stellar masses of $(0.7-4.0) \times 10^{6}$~\msun.  This makes them among the highest redshift young massive star clusters confirmed to date. The inferred stellar mass surface densities also resemble those of local young massive star clusters and globular clusters, suggesting they may be potential globular cluster precursors.

The search and characterization of the (high mass tail) stellar cluster population at high redshift will be soon performed with JWST/NIRCam \citep[][]{NIRCAM} imaging in the lensed fields, allowing us to extend the study presented here to redder wavelengths, until $9000$\AA~rest-frame at z=4, or $5000$\AA~rest-frame up to redshift $z\sim 9$. This highlights the potential of JWST to directly witness the formation of globular clusters during and/or immediately after the epoch of reionization. 

\begin{deluxetable*}{ccccccccccccc}
\tablenum{1}
\tablecaption{Physical properties of the multiple images of System 3, shown in Figure~\ref{fig:pano}. \label{tab:summary}}
\tablewidth{0pt}
\tablehead{
\colhead{ID} & \colhead{M$_{\rm 1600}$} & \colhead{M$_{\rm 4000}$} & \colhead{Mass} & \colhead{Age} & \colhead{SFR} & \colhead{R$_{\rm eff}^{1600 \rm \AA}$} & \colhead{R$_{\rm eff}^{4000 \rm \AA}$} & \colhead{$\Pi$} & \colhead{sSFR} & \colhead{$\Sigma_{Mass}$} &\colhead{$\mu_{\rm Tot}$} & \colhead{$\mu_{\rm Tang}$} \\
\colhead{} & \colhead{HST} & \colhead{JW} & \colhead{[$\times 10^{6}$\msun]} & \colhead{[Myr]} & \colhead{[\msunyr]} & \colhead{[pc]} & \colhead{[pc]} & \colhead{} & \colhead{[Gyr$^{-1}$]}  & \colhead{[\msun~pc$^{-2}$]} & \colhead{}  & \colhead{} 
}
\decimalcolnumbers
\startdata
3.1a  &  -15.09  &  -16.19  &  $4.1^{+0.9}_{-3.3}$  &  $79_{-74}^{+21}$  &  $0.05_{-0.01}^{+0.10}$  &  13.8  &  15.2  &  $21.1_{-14.4}^{+38.8}$  &  12.5  & 1602 &  49.0  &  30.3\\ 
3.2a  &  -14.16  &  -14.56  &  $0.7^{+0.2}_{-0.2}$  &  $22_{-10}^{+9}$  &  $0.03_{-0.00}^{+0.01}$  &  3.2  &  3.8  &  $21.2_{-14.4}^{+39.0}$  &  42.3  & 3979 &  105.1  &  65.2\\ 
3.3a  &  -14.86  &  -15.01  &  $0.9^{+2.2}_{-0.5}$  &  $25_{-16}^{+101}$  &  $0.04_{-0.01}^{+0.02}$  &  11.4  &  11.4  &  $4.3_{-2.9}^{+7.8}$  &  37.7  & 650 &  33.3  &  20.2\\ 
3.1b  &  -15.14  &  -16.09  &  $3.6^{+1.6}_{-2.8}$  &  $63_{-58}^{+37}$  &  $0.06_{-0.01}^{+0.11}$  &  15.3  &  14.5  &  $13.5_{-9.2}^{+24.8}$  &  15.9  & 1547 &  46.8  &  27.3\\ 
3.2b  &  -13.79  &  -14.09  &  $0.5^{+0.2}_{-0.2}$  &  $25_{-13}^{+15}$  &  $0.02_{-0.00}^{+0.01}$  &  3.0  &  4.0  &  $22.7_{-15.5}^{+41.7}$  &  37.4  & 2772 &  102.7  &  62.2\\ 
3.3b  &  -15.22  &  -15.10  &  $1.2^{+1.9}_{-0.5}$  &  $22_{-12}^{+57}$  &  $0.05_{-0.01}^{+0.02}$  &  10.9  &  13.1  &  $4.5_{-3.1}^{+8.3}$  &  42.6  & 610 &  31.4  &  19.1\\ 
3c  &  -18.17  &  -19.32  &  $164.8^{+42.3}_{-125.5}$  &  $126_{-108}^{+33}$  &  $1.47_{-0.25}^{+0.85}$  &  279.4  &  270.9  &  $2.3_{-1.6}^{+4.3}$  &  8.9  & 201 &  3.8  &  2.7\\ 
\hline
\hline
\enddata
\tablecomments{Column (1) lists the IDs of the sources; from top to bottom the coordinates are (3.5893762, -30.3938584) - 
(3.5892221, -30.3938434) - (3.5894880, -30.3938704) - (3.5888077, -30.3937976) - (3.5889706, -30.3938143) - (3.5886425, -30.3937845) - (3.5766318, -30.4018181);
(2) and (3) show the absolute magnitudes at 1600\AA~and  4000\AA; (4), (5) and (6) report the stellar mass, age and SFR derived from the SED fitting; the effective radii at 1600\AA~and 4000\AA~are shown in columns (7) and (8); the dynamical age in units of crossing time ($\Pi$), the specific star formation rate (sSFR) and the stellar mass surface density in columns (9), (10) and (11), respectively; total and tangential magnifications are listed in columns (12) and (13). Uncertainties on the radii are of the order of 50\%, while magnifications have a statistical 3-sigma error of 30\% \citep[][]{bergamini22_a2744}. }
\end{deluxetable*}

\begin{acknowledgments}
This work is based on observations made with the NASA/ESA/CSA James Webb Space Telescope. The data were obtained from the Mikulski Archive for Space Telescopes at the Space Telescope Science Institute, which is operated by the Association of Universities for Research in Astronomy, Inc., under NASA contract NAS 5-03127 for JWST. These observations are associated with program JWST-ERS-1342. We acknowledge financial support from NASA through grant JWST-ERS-1324. KG and TN acknowledge support from Australian Research Council Laureate Fellowship FL180100060. MB acknowledges support by the Slovenian national research agency ARRS through grant N1-0238.
We acknowledge financial contributions by PRIN-MIUR 2017WSCC32 "Zooming into dark matter and proto-galaxies with massive lensing clusters" (P.I.: P.Rosati) and 2020SKSTHZ, INAF ``main-stream'' 1.05.01.86.20: "Deep and wide view of galaxy clusters (P.I.: M. Nonino)" and INAF ``main-stream'' 1.05.01.86.31 "The deepest view of high-redshift galaxies and globular cluster precursors in the early Universe" (P.I.: E. Vanzella).

\end{acknowledgments}

%

\vspace{5mm}
\facilities{JWST(NIRISS), HST(ACS/WFC3), VLT(MUSE)}





\appendix
\label{appA}
\section{JWST/NIRISS photometry probing the rest-optical at z=4}

\begin{figure}[ht!]
\plotone{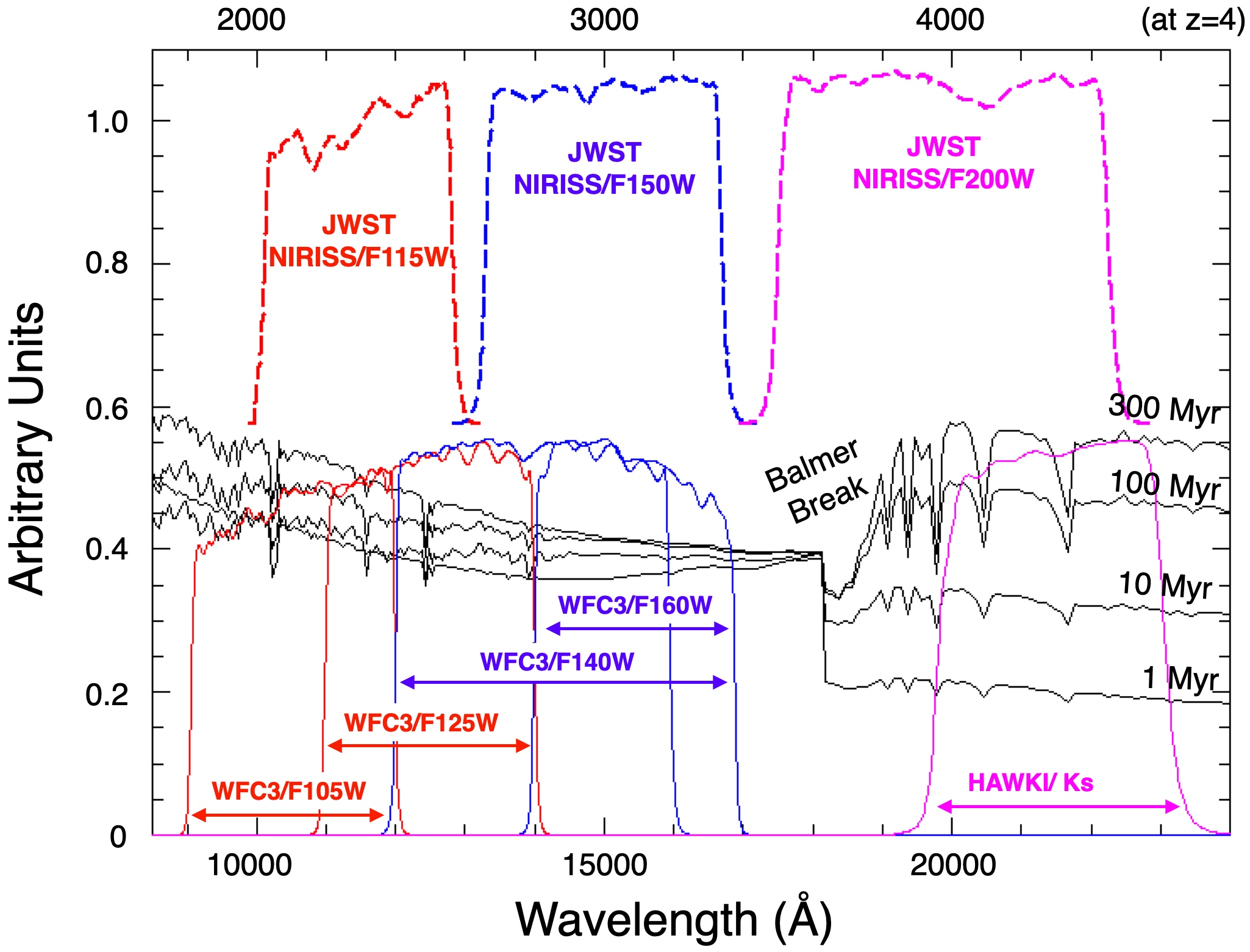}
\caption{The collection of JWST/NIRISS, HST ACS and WFC3 filters, along with VLT/HAWKI Ks-band are shown with superimposed four examples of a z=4 star-forming templates extracted from Starburst99 library \citep[][]{leitherer14}. JWST/NIRISS F200W and VLT/HAWKI Ks bands probe the Blamer break at z=4. HST F105W and F125W filters bracket NIRISS/F115W; similarly, HST/F140W and HST/F160W bracket NIRISS/F150W. \label{filters}}
\end{figure}

Figure~\ref{filters} shows the three NIRISS filters F115W, F150W and F200W compared to the available Hubble Frontier Field ones, HST/WFC3 F105W, F125W, F140W and F160W, along with the ground-based VLT/HAWKI Ks-band. For visualization-only-purpose, four spectral templates placed at z=3.9805 showing the Balmer break at ages of 1, 10, 100, 300 Myr are also superimposed (extracted from Starburst99 library, \citealt{leitherer14}).
The NIRISS/F200W filter probes the Balmer break at an unprecedented angular resolution. 
The photometry of each knot was extracted from PSF-matched HST images \citep[][]{merlin16} adopting circular apertures of $0.27''$ diameter (as we did recently in a study of compact high-z clumps by \citealt{Mestric22}).
Given the superposition of HST and NIRISS filters in the wavelength range $1-1.7$~$\mu m$ and the relatively flat ultraviolet slope of the sources (Figure~\ref{fig:galfit}), we anchor the NIRISS/F150W magnitude to HST WFC3/F140W and F160W (by imposing the NIRISS/F150W flux to match the mean flux computed among the two HST bands bracketing the NIRISS/F150W). We then apply the same shift to NIRISS/F115W and F200W. This makes HST and JWST/NIRISS colors consistent. 
The colors of the SED in the wavelength range $1-2.5\mu m$ are consistent to those observed previously for the full arclet \citep[][]{castellano16}, though here the JWST/NIRISS angular resolution allows us to probe unprecedented spatial scales in the near-infrared.

The normalization of the SED of each knot is subsequently calculated from the total magnitude inferred with {\rm Galfit} in the reference band, HST/F814W (such normalization introduces an additional uncertainty of about 30\%). SEDs were then injected into the SED-fitting code which provided the physical quantities. The stellar mass and the star formation rate  have been de-lensed (divided by $\mu_{Tot}$) to infer the intrinsic properties reported in Table~\ref{tab:summary}.

\bibliography{sample631}{}
\bibliographystyle{aasjournal}



\end{document}